\documentclass[prd,aps,twocolumn,showpacs,superscriptaddress,floatfix]{revtex4-1}
\usepackage{bm}
\usepackage{amsmath}
\usepackage{txfonts}
\usepackage{graphicx}
\usepackage{dcolumn}
\usepackage{bm}
\usepackage{color}
\usepackage{booktabs}
\usepackage{multirow}
\def\comment#1{}

\def\slashchar#1{\setbox0=\hbox{$#1$}           
   \dimen0=\wd0                                 
   \setbox1=\hbox{/} \dimen1=\wd1               
   \ifdim\dimen0>\dimen1                        
      \rlap{\hbox to \dimen0{\hfil/\hfil}}      
      #1                                        
   \else                                        
      \rlap{\hbox to \dimen1{\hfil$#1$\hfil}}   
      /                                         
   \fi}                                         %

\begin{document}

\title{A Dramatically Growing Shear Rigidity Length Scale in a Supercooled Glass Former ($NiZr_2$)}

\author{Nicholas B. Weingartner}
\affiliation{Institute of Material Science and Engineering, Washington University, St. Louis, MO 63130, U.S.A.}
\affiliation{Department of Physics, Washington University, St. Louis,
	MO 63130, U.S.A.}
\email{weingartner.n.b@wustl.edu}
\author{Ryan Soklaski}
\affiliation{Institute of Material Science and Engineering, Washington University, St. Louis, MO 63130, U.S.A.}
\affiliation{Department of Physics, Washington University, St. Louis,
	MO 63130, U.S.A.}
\author{K. F. Kelton}
\affiliation{Institute of Material Science and Engineering, Washington University, St. Louis, MO 63130, U.S.A.}
\affiliation{Department of Physics, Washington University, St. Louis,
	MO 63130, U.S.A.}
\author{Zohar Nussinov}
\affiliation{Institute of Material Science and Engineering, Washington University, St. Louis, MO 63130, U.S.A.}
\affiliation{Department of Physics, Washington University, St. Louis,
	MO 63130, U.S.A.}
\affiliation{Department of Condensed Matter Physics, Weizmann Institute of Science, Rehovot 76100, Israel}
\email{zohar@wuphys.wustl.edu}

\date{\today}

\begin{abstract}
Finding a suitably growing length scale that increases in tandem with the immense viscous slowdown of supercooled liquids is an open problem associated with the “glass transition”. Here, we define and demonstrate the existence of one such length scale which may be experimentally verifiable. This is the length scale over which external shear perturbations appreciably penetrate into a liquid as the glass transition is approached. We provide simulation based evidence of its existence, and its growth by at least an order of magnitude, by using molecular dynamics simulations of $NiZr_2$, a good fragile glass former. On the probed timescale, upon approaching the glass transition temperature,$T_g$, from above, this length scale, $\xi$, is also shown to be consistent with Ising-like scaling, $\xi \propto \left({\frac{T-T_g}{T_g}}\right)^{-\nu}$, with $\nu \approx$ 0.7. Furthermore, we demonstrate the possible scaling of $\xi$ about the temperature at which super-Arrhenius growth of viscosity, and a marked growth of the penetration depth, sets in.  Our simulation results suggest that upon supercooling,  marked initial increase of the shear penetration depth in fluids may occur in tandem with the breakdown of the Stokes-Einstein relation.
\end{abstract}


\maketitle

{\it{Introduction.}}
When a liquid is cooled sufficiently quickly to temperatures well below its melting temperature, nucleation is avoided and the transition to the crystalline state, possessing both extended long-range structural order and absolute minimum free energy, is bypassed. A liquid maintained beneath its melting temperature exists in metastable equilibrium, and is said to be supercooled. A supercooled liquid lacks the long range structural order characterstic of the underlying crystalline ground state, instead maintaining the amorphous atomic arrangement typical of a liquid. As the temperature of the supercooled liquid is lowered further, the viscosity (and relaxation time) increases dramatically, by up to some 14 decades over a temperature range as small as 100 K. Eventually, a temperature, $T_g$, is reached at which the viscosity (and hence, relaxation time) is so large ($>$$10^{13}$ Poise/100 s) that structual rearrangments cease to take place on any reasonable timescale, and the liquid behaves rigidly in response to fluctuations and perturbations. By definition, the liquid is then out of equilibrium, and this is deemed the glass transition. The 'transition' occuring at $T_g$, is in fact not a thermodynamic transition, but instead a kinetic crossover. There is no thermodynamic driving force (energy saving) associated with $T_g$, and a structural rearrangment and associated symmetry breaking is apparently absent. In addition to the smooth emergence of rigidity at $T_g$, the glass transition is accompanied by a rich phenomenology and wide range of interesting features that cannot be enumerated here, but are discussed in a variety of exceptional reviews, e.g. \cite{1,2,3,4,5,6}.

The two most puzzling aspects of the glass transition are the onset of structural rigidity without apparent long range structural order (with associated 
long-time, non-zero shear modulus), and the dramatic, faster than Arrhenius increase of the viscosity/relaxation time found in the so-called ‘fragile’ glass formers \cite{7,8}. Both features seem to call for, and likely require, the existence of a growing length scale, intimately connected to the propagation of some form of ‘amorphous order’ or increasingly cohesive, extended network. In fact, while the notion of a growing activation energy barrier is clearly tied to cooperative motion, even more fundamentally, simple intuitive reasoning suggests that a dramatically growing (or diverging) timescale to relaxation should be coupled to a similarly increasing (and possibly diverging) length scale. Recently, rigorous bounds mandating the existence of a concomitant growth of spatial length scale with relaxation time have been proven to exist \cite{9}. 

The notion of a growing length scale underlying the dynamic slowdown of the glass transition is a principle feature of most theories of glass formation, such as that of Adam and Gibbs, Random First Order Transitions, Mode-Coupling Theories, Kinetically Constrained Models, and others \cite{1,2,3,4,5,6,10,11,12,13,14,15,16,17,18}. Some of these theories predict an underlying phase transition at a temperature below $T_g$, with the glass transition serving as a kinetic "ghost" preceding the actual thermodynamic change. Others posit that there is no true thermodynamic transition besides the melting/freezing transition, and that the length scale corresponds to a geometrically arrested structural ordering which is still capable of bringing about rigidity. As such, the quest to find physical, verifiable, and suitably increasing length scales has been underway for decades. Many proposals for appropriate length scales have been made including those associated with liquid-like defects, the lowest eigenvalues of the relevant Hessian matrix for a system, various point-to-set lengths, elasticity lengths \cite{19,20,21,22,23,24,25,26,27,28,29,30,31,32,33,34}, and  dynamical heterogeneity lengths \cite{35,36,37,38,39,40,41,42,43} and computer vision methods to ascertain both static and dynamic length scales \cite{44}. "Hybrid" correlation length scales, that have mixed static/dyanmical charactersitics, have also been found, as in \cite{Glotzer}. Each of the previously proposed length scales is exceptionally interesting in their own right (and perhaps many can eventually be found arise from the same underlying mechanism), but many display the same drawbacks. Previous numerical and experimental work has shown that these length scales evade experimental verification, and/or do not display an exceptional growth upon approach to $T_g$. For instance, in the case of the lengthscale investigated in \cite{Glotzer}, the behavior of the lengthscale and underlying physics bears a passing resemblance to the investigation done in this work, but that lengthscale, which arises in response to internal perturbations associated with thermal fluctuations, requires knowledge of individual particle displacements, making it difficult to experimentally detect. It is natural to expect that propagating amorphous order should be able to be revealed experimentally, and our proposed correlation length has the benefit of being readily measurable with methods beyond scattering experiments.

It is also worth noting that in \cite{Busch}, the authors found evidence of a {\textbf{decreasing}} correlation length upon cooling toward the glass transition temperature, $T_{g}$, in kinetically strong glassforming liquids. This was found to be indicative of a lambda transition in the vicinity of the glass transition. It was further suggested that for kinetically fragile liquids, the behavior would be the opposite, with the correlation length increasing upon approach to $T_{g}$. While this general behavior is consistent with our findings, we found no evidence suggestive of a lambda transition in this system, and further, the correlation length investigated in \cite{Busch}, was purely dynamical in nature. 

In light of the above discussion, and based upon suggestions made in previous theoretical work \cite{44,45}, we perform molecular dynamics simulations of $NiZr_2$, an excellent representative of a fragile glass \cite{46}. {\it{We provide evidence for the rapid increase of the shear penetration depth}}, defined as the length over which a supercooled liquid rigidly responds to externally imposed forces. We find that the near divergence of the penetration depth as the system becomes glassy is not far off the mark of Ising-like scaling.

{\it{The Shear Penetration Depth}.} 

\begin{figure}
\includegraphics[width=1 \columnwidth, height= .4 \textheight, keepaspectratio]{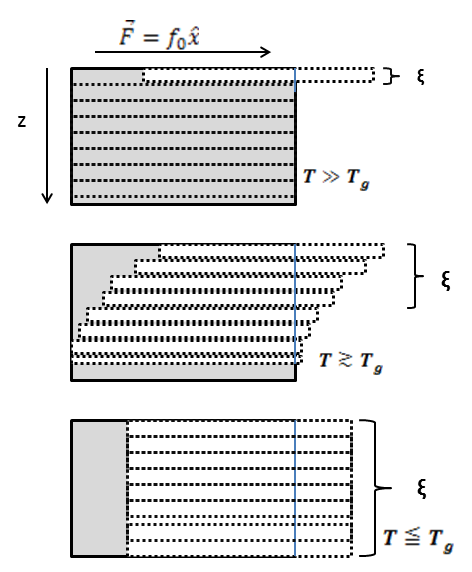}
\caption{(Color Online) Representation of the proposed response of general supercooled fluid systems. The solid lines represent the original box shape before perturbation. The dashed regions represent the successive layers that respond to the perturbation at temperatures above and around Tg. At high temperatures only the layers experiencing the external stress move appreciably, but as temperature, T, is lowered and the cooperativity becomes pronounced, the perturbation is transmitted deeper into the material, reflecting increasing rigidity. Note that the extent to which the layers move as depicted, have been greatly exaggerated for clarity.}
\label{Depth.}
\end{figure}

In ordinary critical phenomena, the correlation length scale is defined as the typical spatial extent of a fluctuation of the thermodynamically relevant order parameter. It can also be interpreted as the average length over which a perturbation by the appropriate conjugate "generalized force"  will appreciably propagate. For example, in the Ising Model, the correlation length corresponds to fluctuations in the typical size of magnetic domains (the order pameter is the magnetization, $\vec{M}$), and the also corresponds to the distance over which an applied magnetic field, $\vec{\mathcal{B}}$ (the conjugate force) will influence the system. As it is known that in crstalline solids, the rigidity is due to long range order, we can apply this idea to the glass transition problem. In this case the ordering should be short range at temperatures just below melting, and grow as temperature as lowered. We can quantify this by subjecting the liquid to a shear perturbation on the boundary, and tracking how it penetrates the liquid transverse to the applied stress. Liquids, by definition, are capable of rearrangement to dissipate shear stress; a shear force applied to the top of a liquid will only propagate appreciably through a finite number of layers below the perturbation before fully decaying. Intuitively, one would expect that a high temperature liquid, having relatively low viscosity, would respond to the external force in a manner such that only the forced layer experiences a substantial displacement relative to the opposite boundary. As the temperature is lowered and the viscosity increases, one expects an associated increase in the liquid’s effective, short-lived rigidity. This, we argue, corresponds with increasing structural ordering and kinetic cohesion of network-like structures in the liquid. At moderate supercooling, then, one expects a deeper penetration of shear perturbations and associated displacements sustained by layers of the liquid that are increasingly distant from the applied perturbation. As depicted in Figure 1, the shear is applied to the top layer of the simulation box, transverse to the vertical (z) direction. The penetration depth is defined as the distance (along the z axis) up to which appreciable effects of shear are observed. At $T_g$ when solid-like rigidity has set in, one expects that the whole block of material will roughly slide together, such that the penetration depth is the length of the material. This is consistent with results that show the continuous emergence of a finite shear modulus for temperatures below $T_g$ \cite{47}. This process is pictured, schematically, in Figure (\ref{Depth.}). Ultimately this length scale is agnostic to the specific type of structural ordering, but can be related to cooperativity and the idea of a divergent correlation length in ordinary critical theory. 

{\it{Models and Methods.}}
Molecular dynamics (MD) simulations were employed using the LAMMPS package \cite{48}. The atoms in the simulation evolved under the influence of a semi-empirical Finnis-Sinclair type Embedded Atom Model potential created by Mendelv et al \cite{49} with periodic boundary conditions. The parameters and coefficients associated with the potential were fitted using X-ray diffraction data as well as enthalpy of mixing values, and volume measurements in the liquid state. This potential has been shown to excellently reproduce both the high temperature liquid as well as glassy states of $NiZr_2$ \cite{49}.

The simulations were run in the NPT ensemble with N=5000 atoms and a target external pressure of P=0. Thermostatting and barostatting were employed using a Nose-Hoover thermostat, and barostat respectively, and the velocity-verlet algorithm was utilized to integrate the equations of motion. A 5fs timestep was employed. The initial configurations were generated randomly and the atoms were then allowed to melt and evolve naturally for 0.25 ns at a temperature of 2200 K to allow for equilibration. The system was then quenched to various target temperatures ranging from 300 K up to 1900 K using a quench rate of $\mathcal{Q}=10^{13}$ K/s. After the quench, the system was allowed to evolve unperturbed for an additional 0.1 ns. The process was repeated, starting from independent initial configurations, for $T_g$ as well as all sampled temperatures below $T_g$ and some representative temperatures above. 

\begin{figure}
\includegraphics[width=1 \columnwidth, height=.6 \textheight, keepaspectratio]{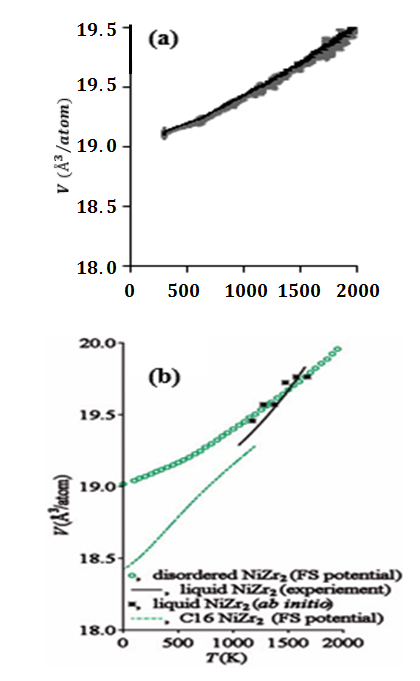}
\caption{(Color Online). Panel (a): Specific volume as a function of temperature for our simulated system. Panel (b): Specific volume as a function of temperature. Reproduced from \cite{49}.}
\label{Volume.}
\end{figure}

As the glass transition is a kinetic phenomenon without a thermodynamic driving force, the glass transition temperature $T_g$ is not a constant, and weakly depends on cooling rate, and external timescale. Therefore, one has to be careful in identifying its precise location. Typically, various thermodynamic parameters show a crossover at the glass transition, associated with falling out of equilbrium (the system loses its translational degrees of freedom on the timescale of observation). One such property that shows a change in behavior at the glass transition is the volume. The temperature dependence of the volume shows a "kink" at the glass transition temperature, $T_g$ (the thermal expansion coefficient, $\alpha \equiv \frac{1}{V} \frac{\partial V}{\partial T}$, has a discontinuity), providing an efficient way to determine $T_g$. The temperature dependence of the volume of our system during a quench to 300 K is depicted in panel (a) of Figure (\ref{Volume.}). There appears to be a subtle kink in the vicinity of $T \approx 700 K$. This is in good agreement with the results in panel (b) of the same figure, which was produced by the author of the potential in \cite{49}, as well as previous numerical work performed under similar protocols \cite{50}. 

\begin{figure}
\includegraphics[width=1 \columnwidth, height= .5 \textheight, keepaspectratio]{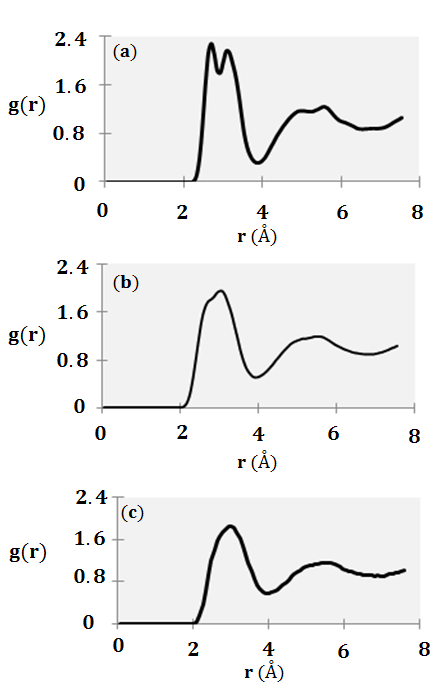}
\caption{(Color Online). Radial Distribution Function at a temperature of (a): T=300 K $<T_g$, (b): moderate supercooling with T=1100 K, and (c): above the melting temperature at T=1500 K.}
\label{Distribution.}
\end{figure}
In order to assess that our system is behaving as expected before applying shear stresses, we examine the behavior of the radial distribution functions at various temperatures. These results are shown in Figure (\ref{Distribution.}). It is clear that system behaves as expected as the glass transition is approached, and in comparison with \cite{49}, we see that the radial distribution functions (RDFs) measured in this work retain the overall shape and placement of the peaks. The height of the first peak (and behavior of the splitting of the first peak), however, is slightly different from those in \cite{49} and we attribute this to the discrepancy in quench rates. 

We modeled the external shear stress by defining a 4 angstrom-thick layer at the top of our simulation box, and applying an external force in the x direction on the atoms in this layer. In order to avoid fracturing in such a small system size at low temperatures, a force value of only 0.2 eV/A was used. A stronger force would be expected to make the effects more dramatic, but system size limitations did not allow for higher values. The force was left active for 100 timesteps to attempt to approximate an impulsive “kick” at the top of the box. No external forces were applied to the bottom of the box. After an observation time of $\tau_o$=16,000 timesteps, displacement data was extracted.

{\it{Measurement Results.}}
\begin{figure*}
\centering
\includegraphics[width= 1.8 \columnwidth, height= .5 \textheight, keepaspectratio]{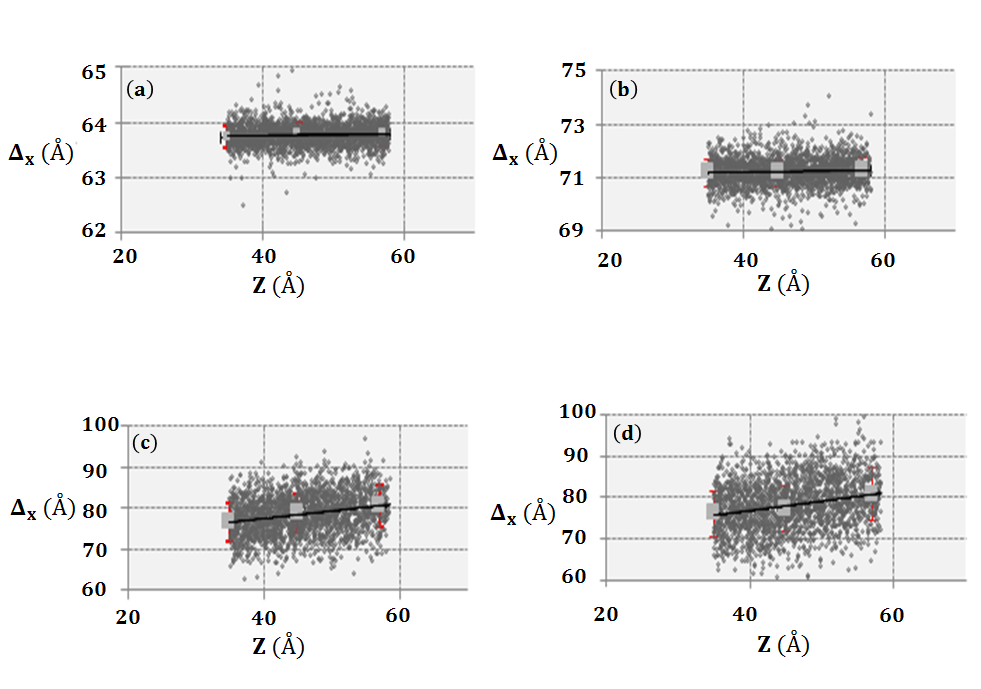}
\caption{(Color Online). Typical Data at four [(a): 300 K, (b): 700 K, (c): 1300 K, (d): 1500 K] representative temperatures both above and below $T_g$ ($\approx$ 700 K). The red lines correspond to the standard deviation at the two boundaries and center of the material. It is noteworthy that they are significantly tighter than the data seems to suggest at this level of zoom. The black lines are the lines of fit from which the slope is extracted to define the length scale. Note the dramatically changing behavior as T is lowered.}
\label{Displacement.}
\end{figure*}

\begin{figure}
\includegraphics[width=1 \columnwidth, height=.45 \textheight, keepaspectratio]{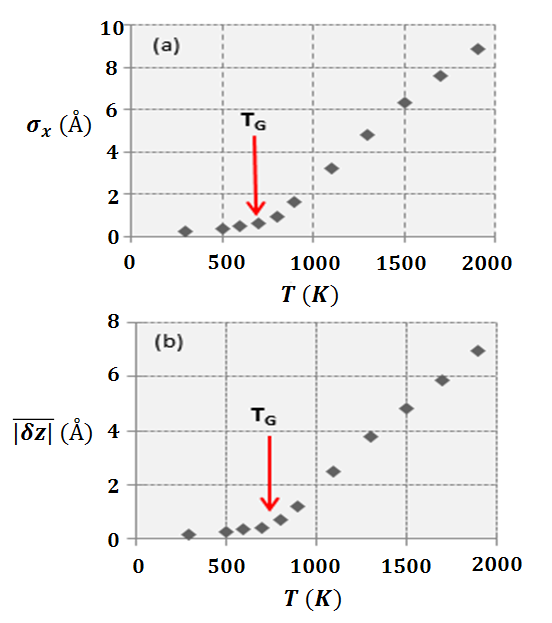}
\caption{(Color Online). Panel (a): Plot of standard deviation of displacement in direction (x-axis) of applied shear as a function of temperature. Panel (b): Average displacement in height dimension (z-dimension) as a function of the temperature.}
\label{Noise.}
\end{figure}

To quantify the depth of penetration of the shear stress, we plot the displacement of each atom in the shear direction (x-direction) versus its position in the transverse height dimension (z-direction). The displacement represents the net movement in the shear direction from the timestep before the external shear stress was applied, up to the observation time $\tau_o$ (as described in methods section). The height of each particle corresponds to the vertical layer it is in at the observation time. Due to the periodic boundaries, only atoms in the layers from $z=\frac{L}{2}$ to $z=L$ at the observation time were considered. Figure (\ref{Displacement.}) shows displacement data for four representative temperatures, i.) deep in the glassy phase (300K), ii.) at $T_g$ (700K), iii.) in the moderately supercooled regime (1300K), and iv.) above $T_{melt}$ (1500K). Thermal effects tended to produce large motions in the height dimension at temperatures above $T_{melt}$, but the effects were not large enough to wash out the effect of shear penetration except at very high temperatures (1700-1900 K) . Figure (\ref{Noise.}) serves to quantify the impact of thermal noise. In panel (a), the standard deviation of displacement (at the observation time) in the direction of applied shear is plotted as a function of the temperature. As expected, the thermal noise decreases with decreasing temperatures becoming very small as $T_g$ is approached. In panel (b), the average magnitude of the particle displacements in the height (z) direction is plotted as a function of temperature. It is clear from panel (b) that large scale thermal motion may play a role in "washing out" the shear penetration depth at the highest temperatures measured. 

\begin{figure}
\includegraphics[width=1 \columnwidth, height=.35 \textheight, keepaspectratio]{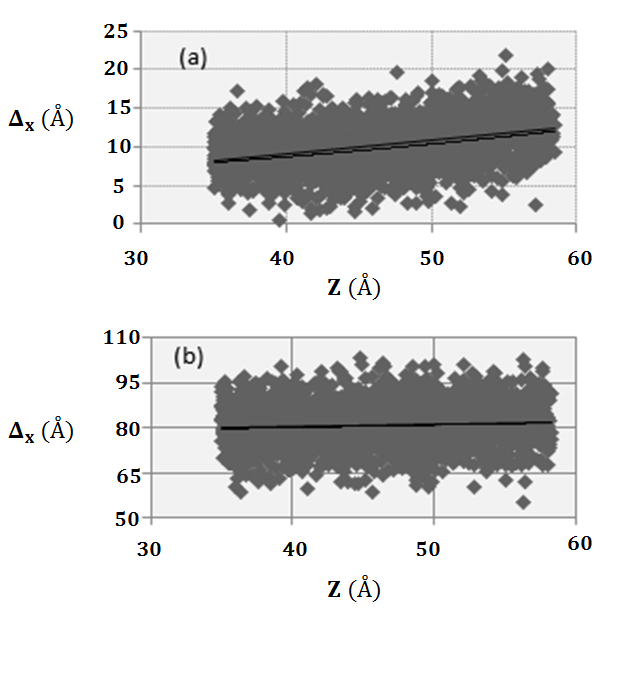}
\caption{(Color Online) Displacement data for a random configuration at 1700 K. Panel (b) is the displacement for the observation time (16,000 timesteps), panel (a) is the data at an earlier time (3000 timesteps after the shear is turned off). As panel (b) shows, at high temperature the shear induced displacements appear to be far smaller at the standard observation time used in this work. Nevertheless, at earlier times, as seen in panel (a), the displacements are much more noticeable.}
\label{Temporal.}
\end{figure}

The general response function, $\mathcal{R}$, of our system to the externally imposed shear is a  function of distance in the z direction to the imposed shear, temperature T, and observation time $\tau_o$;
\begin{eqnarray}
\label{Response}
\mathcal{R}=\mathcal{R}(z,T,\tau_o).
\end{eqnarray}
In this work, we chose a constant observation time, $\tau_o$, and varied the temperature to ascertain the penetration depth along the z axis. The penetration depth of the applied shear ultimately has some value depending on the temperature. However, our ability to extract the exact value depends on the observation time chosen. For sufficiently short observation times, the effects of the externally applied shear cannot penetrate the system at the lowest temperatures (near and below $T_g$). Therefore, the observation time has to be sufficiently long to capture the effect at low temperatures. As shown in Figure (\ref{Temporal.}), for very long observation times at high temperatues, the effects of the external shear will be nill. Therefore, using an observation time which is very long would lead one to conclude a much deeper penetration depth at high temperatures (see Figure \ref{Temporal.}). Hence, choosing an appropriate observation time is important. For a couple representative temperatures we investigated the impact of observation time. In each case the duration for which the shear force was applied was constant; for this work we wanted to maintain an approximation of an impulsive kick to the system. Oscillatory shears have been discussed elsewhere.  It was observed that at the lowest temperatures (T<$T_A$), the results showed little change when observation times were changed by factors of two. For high temperatures, the observation time plays a more noticeable effect. Afte investigation, we found that the observation time employed in this work was sufficient to capture the low temperature effect of the penetration, while not losing the high temperature impact except at the hightest temperatures studied. 

Typically, one would expect the displacement response to decay exponentially with depth.  As pictured in Figure (\ref{Displacement.}), at these system sizes, the shear-induced displacement, while not very large, is still quite noticeable. As the displacement was not extremely large, we applied a linear regression to the data rather than an exponential one.  These fits are sufficient to quantify the penetration depth. Fits to the data were of the form,
\begin{eqnarray}
\label{Linear}
\delta_x(z,\tau_o)=m*z+\delta_0
\end{eqnarray}
where we denote by $\delta$, the displacement in the shear direction (x-direction) and z the height of the layer (both measured in angstroms). We define the rigidity length scale (the penetration depth) as
\begin{eqnarray}
\label{Depth}
\xi \equiv \frac{1}{m}
\end{eqnarray}
where “m” is the slope in Eq. (\ref{Linear}). For the temperatures noted above, the value of m was averaged over multiple runs, and this average was used in Eq. (\ref{Depth}) for these temperatures. 

\begin{figure}
\includegraphics[width=1 \columnwidth, height=.35 \textheight, keepaspectratio]{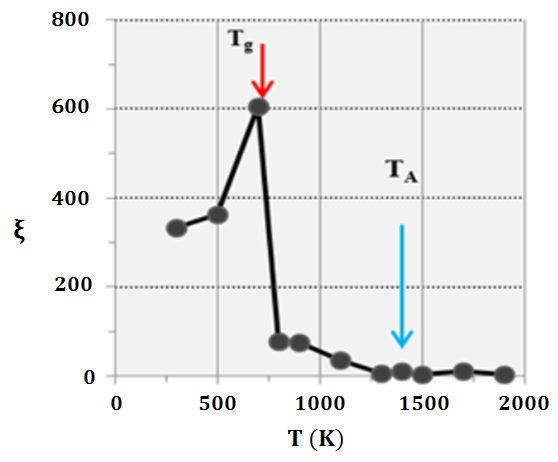}
\caption{(Color Online). Plot of the length scale, $\xi$ versus temperature. All temperatures below $T_g$ were averaged over multiple independent runs, as were select, representative temperatures above $T_g$.}
\label{Lengthscale.}
\end{figure}
The temperature dependence of the length scale ($\xi$) is shown in Figure (\ref{Lengthscale.}). A dramatic growth of $\xi$ with decreasing temperature is evident. The first notable penetration of the shear, beyond the layer to which the force was applied, occurs at a temperature marked $T_A (\approx 2 T_g)$ \cite{51}. Below this temperature super-Arrhenius growth of the viscosity may be anticipated based upon collective effects \cite{51,52,53,54}. The sudden, monotonic increase of $\xi$ at temperatures below $T_A$, provides direct support to earlier numerical studies, which found that metallic liquids begin to develop solid-like features once they are cooled below $T_A$ \cite{51,52}. These solid-like features include the breakdown of the Stokes-Einstein relationship, exponential stretching of the relaxation functions, and the onset of cooperative structural rearrangements during the liquid’s relaxation process \cite{1}. Indeed, $T_A$ does appear to serve as a crossover temperature below which the liquid begins to exhibit a substantial rigid response to external forces, though it is a local and transient response. Putting all this together, our simulation results allow us to predict that marked growth of the shear penetration depth may commence at the same temperature as the breakdown of the Stokes-Einstein relation in real supercooled liquids. The penetration depth increases rapidly as the liquid is supercooled toward the glass transition temperature, $T_g$. Below the latter temperature, the material is “glassy” and exhibits structural rigidity on all practical timescales, such that shear perturbations propagate the length of the material and appear to diverge. 

\begin{figure}
\includegraphics[width=1 \columnwidth, height=.45 \textheight, keepaspectratio]{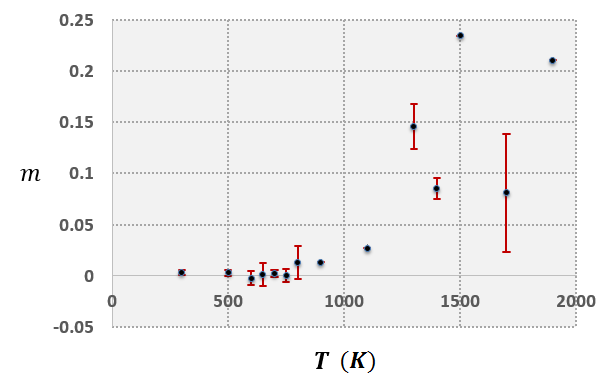}
\caption{(Color Online). Plot of the average value of the slopes, m, at each of the measured temperatures along with their associated standard deviations $\sigma$ (when multiple runs were performed).}
\label{Signal_Noise.}
\end{figure}
As discussed previously, we performed multiple (typically six) independent measurements at all temperatures $T \leq T_g \approx 700K$, as well as most of the representative temperatures above $T_g$ (300-800, 1300, 1400, 1700K). At each of these temperatures, averaging was done to determine the slope, m, in the fit of Eq. (\ref{Linear}). The apparently periodic nature of $\xi$ below $T_g$ can be attributed to noise due to the lengthscale being essentially divergent to the system size at these temperatures. This point is vividly made in Figure (\ref{Signal_Noise.}), which depicts the average slopes, m, as a function of temperature. The error bars on the points with multiple runs corresponds to the standard deviation in slopes. We see that, at temperatures near and below $T_g$, the combination of the average value and associated error bars lead to m being virtually indistinguishable from zero, consistent with the divergence of the penetration depth beyond the system size at the glass transition. It should also be immediately noticeable that the data points associated with the value of the penetration depth at temperatures  T=600, 650, and 750 K, are not plotted in Figure (\ref{Lengthscale.}). This is because, as seen in Figure (\ref{Signal_Noise.}), the signal to noise ratio ($\frac{\sigma}{m}$) for these points was a factor of three for the data point at 600K and a factor of ten for the data points at 650 and 750 K.  This large relative error is due to the fact that the length scale, $\xi$, is exceedingly large as the corresponding average slope m is very small, and in fact virtually indistinguishable from zero (see Eq. (\ref{Depth})). This corresponds to total penetration of the shear to beyond the system size, and the fluctuations about m=0, are to be expected due to ordinary thermal effects. Because of the large relative error in the aformentioned data points, we removed these data points from Figure (\ref{Lengthscale.}) so as to not mask the overall monotonic increase of $\xi$ with incredible values. It may at first seem concerning that the data point at T=750 K is suggestive of near divergene considering it is above $T_g$. This is, in fact, not an issue, as the value of $T_g$ is not precise, and very likely falls within the T=700 to 750 K range for this system size and quench rate. Also, the close proximitiy to $T_g$ and limitations of resolution at this system size, would lead to the impact of the arrest at the glass transition strongly influencing temperatures asymptotically close to $T_g$. Clearly, Figure (\ref{Signal_Noise.}) serves not only to explain the fluctuations but also reinforces the idea that the penetration depth appears to diverge to the system size in the vicinity of $T_g$, and is perhaps the most consequential and rigorous result in this work.

\begin{figure}
\includegraphics[width= 1 \columnwidth, height=.5 \textheight, keepaspectratio]{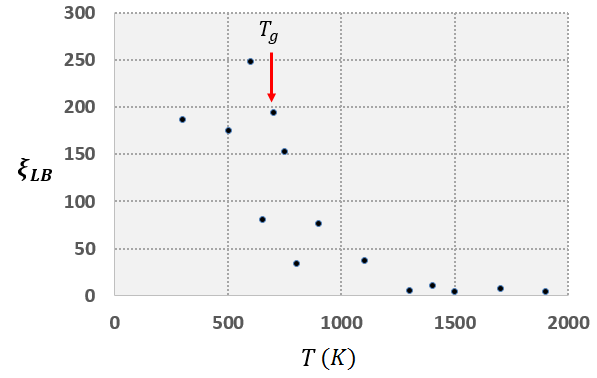}
\caption{(Color Online). Lower bound, $\xi_{LB}$, on the shear penetration depth. See text.}
\label{Lower_Bound.}
\end{figure}
As a lower bound on the shear penetration depth, in Figure (\ref{Lower_Bound.}) we plot 
\begin{eqnarray}
\label{LB}
\xi_{LB} \equiv \frac{1}{m+\sigma}.
\end{eqnarray}
Because this is a lower bound, we can strongly assert, based on our data, that the penetration does indeed show dramatic increase upon supercooling. For temperatures below $T_g$, the length scale is so large and slopes so small, that the observed fluctuations of $\xi$ may be statistical (see Figure \ref{Signal_Noise.}). We also conclude that the length scale becomes, at least, considerably larger than the system size at $T_g$, and may in fact diverge. For the impulsive kick we applied this will remain the case at all longer timescales of observation. However, if one were to apply a {\textbf{static shear}} at the top of the box and left this shear on for a time longer than the relaxation time, then even below $T_g$, the length may not diverge. On all practical timescales, though, it would, not diminishing this length as a natural candidate for the glass transition problem. The precise behavior of the penetration depth as the duration of the static shear stress is varied is an interesting problem, but requires a different type of analysis, and will be addressed in a future work. 

{\it{Scaling Arguments.}}
\begin{figure}
\includegraphics[width=1 \columnwidth, height=.45 \textheight, keepaspectratio]{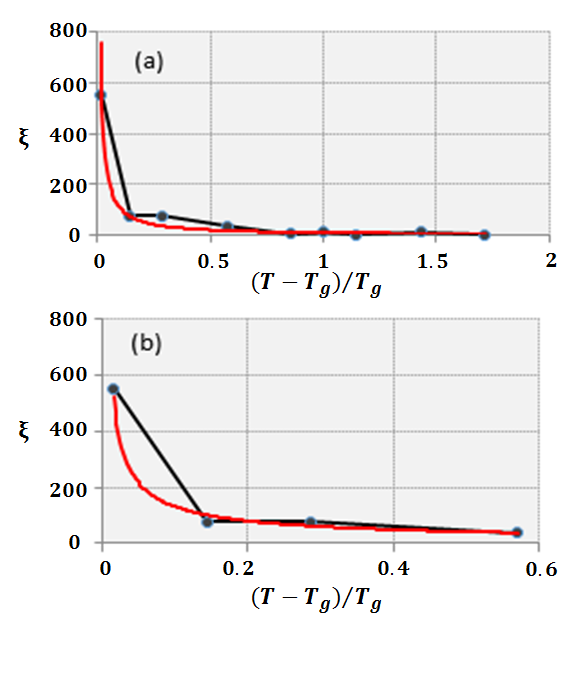}
\caption{(Color Online). Power law fits to the shear penetration depth as a function of reduced temperature (measured relative to glass transition temperature $T_g$). Panel (a): $\xi \propto \left({\frac{T-T_g}{T_g}}\right)^{-1}$. Panel (b): $\xi \propto \left({\frac{T-T_g}{T_g}}\right)^{-0.71}$.}
\label{Scaling.}
\end{figure}
Previous studies have examined the behavior of various proposed length scales in the vicinity of  $T_g$ (or the Vogel-Fulcher-Tammann temperature $T_0$ \cite{7,55,56,57}. In \cite{30}, a diverging length scale associated with “liquid-like” defects at $T_g$ produced an exponent $\nu$=1; This value constitutes an upper bound on the exponents reported in other works. Researchers in \cite{58} found a critical exponent of $\nu$=0.875 for the largest icosahedral cluster size in a model metallic glass former \cite{54}. In both \cite{59}, focusing on inherent structures in a binary Lennard-Jones glass former, and in \cite{60}, by largely studying medium-range bond orientational order in colloidal liquids, scaling analyses gave exponents of $\nu \approx$ 2/3, in rough agreement with a three-dimensional Ising exponent ($\nu$=0.625(1) \cite{61}). We performed a similar scaling analysis of our data, fitting a function of the form
\begin{eqnarray}
\label{Scaling}
\xi \propto \left({\frac{T-T_g}{T_g}}\right)^{-\nu}.
\end{eqnarray}
To represent the value of the length scale at $T_g$ itself, we interpolated the value at T=710K using the line connecting T=800K and T=700K. In addition, the data point at T=750 K, was excluded from this scaling analysis for reasons discussed above. When using the full range of temperatures in applying the power law fit, we extracted an exponent of $\nu$=1. This is depicted in Figure (\ref{Scaling.}), panel (a). Decreasing the temperature range considered in the scaling to only include temperatures very close to $T_g$, caused the value of the exponent $\nu$ to decrease. This is depicted in panel (b) of Figure (\ref{Scaling.}), where a value of $\nu$=0.713 was found. Clearly, as the scaling is applied to a more and more asymptotic region around $T_g$, the value of $\nu$ appears to approach a value consistent with Ising-like scaling. Our value for ν, is, thus, in rough agreement with previously suggested exponents extracted by different means. A similar study conducted in \cite{34} studied the high temperature ($T \geq T_m$) correlations of the anisotropic part of the atomic level stress. In this work the authors found an exponent of $\nu \approx$0.7 for the low temperature extrapolation of the correlations in their two-dimensional system (the high crystallization rate at temperatures below 
$T_m$ thwarted a direct study at low temperatures). Our possible scaling may provide further evidence of a universal nature of the length scale at deep supercooling. As $T_g$ is not a true thermodynamic temperature, it is unclear what this scaling may mean, but it may be suggestive of universality in the glass transition.

\begin{figure}
\includegraphics[width=1 \columnwidth, height=.45 \textheight, keepaspectratio]{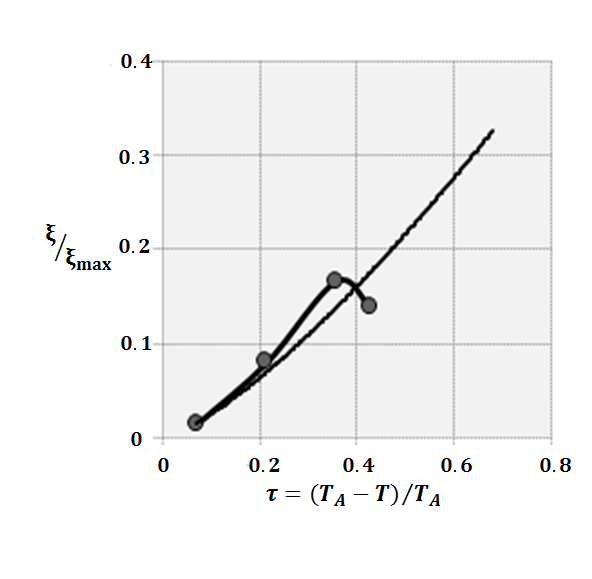}
\caption{(Color Online). Power law scaling, $\xi \propto \left({\frac{T_A-T}{T_A}}\right)^{\nu_A}$, in the asymptotic region below the crossover temperature, $T_A$.}
\label{Ta_Scaling.}
\end{figure}

In [69, 70] theoretical arguments for the scaling of the length scale about the crossover temperature, $T_A$, were provided. It was suggested when asymptotically approaching $T_A$ from below, that a characteristic structural domain size, l,  scaled as 
\begin{eqnarray}
\label{Ta_Scaling}
l={\tau_A}^{\nu_A},
\end{eqnarray}
where $\tau_A \equiv \frac{T_A-T}{T_A}$. We examined the shear penetration depth as a function of the reduced temperature, $\tau_A$, as depicted in Figure (\ref{Ta_Scaling.}). We observed that a power law is a good fit in this region with an exponent, $\nu_A \approx$1.335. If the lowest temperature (highest reduced temperature) point in Figure (\ref{Ta_Scaling.}) is removed, then a value of $\nu_A \approx$1.5 will be obtained instead.

{\it{Conclusion.}}
\begin{figure}
\includegraphics[width=1 \columnwidth, height=.45 \textheight, keepaspectratio]{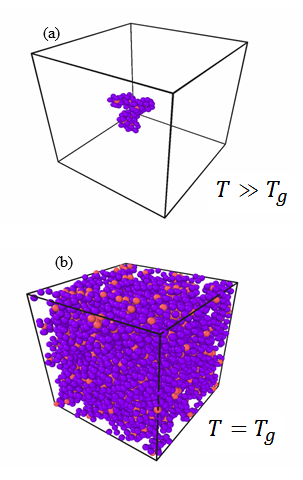}
\caption{(Color Online). Depiction of growing interconnectivity of icosahedral clusters with supercooling in $Cu_{36}Zr_{64}$, a very similar metallic glass former. The Cu atoms are marked red and the Zr by purple. In Panel (a) we show the longest interconnected cluster at 1200 K. Panel (b) shows the longest connected cluster at $T_g$ (800 K for this system). Note that interconnecting icosahedra percolate at T$_g$. (These results are similar to those in \cite{54}.)}
\label{Percolation.}
\end{figure}
We have defined the shear penetration depth as the distance over which supercooled liquids can support shear appreciably. This definition is based on a simple physical picture of liquids as a continuum unable to globally support shear stress. With decreasing temperature the viscous inter-layer forces, rigidity, and the lifetime of connectivity increase. At $T_g$, the glass transition temperature, these quantities mirror those of crystals and the material is solid. This scale has the added benefit that it may be readily experimentally accessible. Results from simulations on $NiZr_2$, a typical ‘fragile’ glass, support this definition. In the current work, we applied, in simulatum, a shear force to the top layer of a $NiZr_2$ system. By measuring the penetration depth, $\xi$, we have demonstrated a dramatically increasing structural scale upon supercooling toward $T_g$. This leads to the conclusion that the shear penetration depth marks a very natural candidate for the structural length scale characterizing glassy dynamics. Furthermore, and of equal importance, the shear penetration depth can be measured experimentally. While it might be practically difficult, in theory, the penetration depth can accessed experimentally in a way that does not rely solely on scattering, and may, therefore, be easier to investigate. This is a major advanatage for this lengthscale, and sets it apart from previously proposed lengths.

We ultimately believe that the shear penetration depth is intimately connected to the structure of the supercooled liquid. It has been shown \cite{52,54,62,63,64,65,66,67} in extensive numerical studies that clusters of locally preferred structural order tend to grow and interconnect as temperature is lowered in supercooled liquids. These clusters locally minimize the relevant free energy and hence are stronger and more stable to fluctuations. Interconnections of the clusters increase in length and lifetimes \cite{52,53,64} upon lowering temperature, and eventually span the system size at the glass transition. This percolation of a structural network, which is depicted in Figure (\ref{Percolation.}) for a ‘cousin’ configuration, is very likely the source of the penetration depth as well as a leading cause of the arrest at the glass transition temperature $T_g$. 

Due to their locally stable nature and tight binding, the clusters resist thermal breakup and lock into a rigid structure forming a force network that can propagate shear. The interlocking and cohesiveness of this network also serves to slow down the dynamics \cite{64}, as sufficiently large thermal fluctuations are needed to break the network and this becomes less likely with lower temperature. In fragile glasses, this network has to form quickly over the temperature range encountered in typical experiments. This is likely due to the largely non-directional binding in fragile glasses which lacks the natural network found in strong covalent liquids.  

In metallic liquids, the network is likely icosahedral \cite{52,54,62,63,65}. In silicates (typical of the strong classification) a natural tetrahedral network with strong bonds and directionality is present. It has been suggested that \cite{65} networks of locally preferred structures tend to form in fragile glasses being either icosahedral or crystal-like at short range. 

The notion that the shear penetration in amorphous solids is due to a system spanning network is a universal one. As discussed a network forms over a narrow range in fragile glasses leading to the super-Arrhenius increase of viscosity and causing the rigidity. In strong glasses, a tetrahedral network forms at high temperature and becomes increasingly cohesive as the temperature is lowered to $T_g$. Other forms of a system spanning network can also exist. In colloids a frictional or contact network can be created by jamming, and in fact a rigidity length scale has been proposed for these systems \cite{26}. The formation of a contact network has also been shown to occur, albeit short-lived, in some discontinuous shear thickening fluids \cite{66}. This contact network may also play a role strong to fragile crossovers in high pressure thermal glasses. 

Based on the above discussion, it is clear that a shear penetration length scale can be quite naturally extended to many, if not most, glassy systems. This leads naturally to the connection between slowing down and network formation. The fact that our length scale begins to grow substantially, only when supercooled beneath $T_A$, further suggests a structural origin for the glass transition.

{\it{Acknowledgements}}.  We would like to thank Bo Sun, Sadegh Vaezi, Matt Blodgett, Li Yang, and Vy Tran for their input, and thoughtful discussion. NW and ZN were supported by the NSF DMR-1411229. ZN thanks the Feinberg foundation visiting faculty program at Weizmann Institute. RS was partially supported by NSF (DMR 1207141) KFK was partially supported by NSF (DMR 12-06707) and NASA (NNX10AU19G).

\end{document}